\documentclass{pasj00}

\newcommand{\EHM}[1]{\mbox{$\times10^{#1}$}}
\newcommand{\pmHM}[2]{\mbox{$^{#1}_{#2}$}}
\newcommand{\lumiHM}{\mbox{erg~s$^{-1}$}}
\newcommand{\fluxHM}{\mbox{erg~s$^{-1}$~cm$^{-2}$}}
\newcommand{\crHM}{\mbox{c~s$^{-1}$}}
\newcommand{\hessLHM}{\mbox{HESS~J1616$-$508}}
\newcommand{\hessHM}{\mbox{HESS~J1616}}

\begin{document}
\SetRunningHead{H. Matsumoto et al.}{Dark Accelerator HESS~J1616$-$508}
\Received{2006/07/28}
\Accepted{2006/08/16}

\title{Suzaku Observations of HESS~J1616-508: 
Evidence for a Dark Particle Accelerator}



%
 \author{%
   Hironori \textsc{Matsumoto}\altaffilmark{1},
   Masaru \textsc{Ueno}\altaffilmark{2},
   Aya \textsc{Bamba}\altaffilmark{3},
   Yoshiaki \textsc{Hyodo}\altaffilmark{1},\\
   Hideyuki \textsc{Mori}\altaffilmark{1},
   Hideki \textsc{Uchiyama}\altaffilmark{1},
   Takeshi \textsc{Tsuru}\altaffilmark{1},
   Katsuji \textsc{Koyama}\altaffilmark{1},\\
   Jun \textsc{Kataoka}\altaffilmark{2},
   Hideaki \textsc{Katagiri}\altaffilmark{4},
   Tadayuki \textsc{Takahashi}\altaffilmark{5},
   Junko \textsc{Hiraga}\altaffilmark{3},\\
   Shigeo \textsc{Yamauchi}\altaffilmark{6},
   John P. \textsc{Hughes}\altaffilmark{7},
   Atsushi \textsc{Senda}\altaffilmark{3},
   Motohide \textsc{Kokubun}\altaffilmark{8},\\
   Takayoshi \textsc{Kohmura}\altaffilmark{9},
   and
   Frederick S. \textsc{Porter}\altaffilmark{10}
}
 \altaffiltext{1}{Department of Physics, Graduate School of Science, 
Kyoto University, Sakyo-ku, Kyoto 606-8502}
 \email{E-mail (HM) matumoto@cr.scphys.kyoto-u.ac.jp}
 \altaffiltext{2}{Department of Physics, Faculty of Science,
Tokyo Institute of Technology, \\
2-12-1, Meguro-ku, Ohokayama, Tokyo 152-8551}
 \altaffiltext{3}{RIKEN, cosmic radiation group,
2-1, Hirosawa, Wako-shi, Saitama, Japan}
 \altaffiltext{4}{Department of Physics, Graduate School of Science, 
Hiroshima University, \\
1-3-1 Kagamiyama, Higashi-Hiroshima, Hiroshima 739-8526}
 \altaffiltext{5}{Institute of Space and Astronautical Science, 
Japan Aerospace Exploration Agency, \\
3-1-1 Yoshinodai, Sagamihara, Kanagawa 229-8510}
 \altaffiltext{6}{Faculty of Humanities and Social Sciences, Iwate University, 3-18-34 Ueda, Morioka, Iwate 020-8550}
 \altaffiltext{7}{Department of Physics and Astronomy, Rutgers University, \\
136 Frelinghuysen Road, Piscataway, NJ 08854-8019, USA}
 \altaffiltext{8}{Department of Physics, 
University of Tokyo, 7-3-1 Hongo, Bunkyo-ku, Tokyo}
 \altaffiltext{9}{Department of General Education, Kogakuin University, \\
2665-1 Nakano-Cho, Hachioji, Tokyo 192-0015,}
 \altaffiltext{10}{
NASA Goddard Space Flight Center, Laboratory for High Energy Astrophysics, \\
Code 662, Greenbelt, MD 20771.}


\maketitle

\begin{abstract}

We observed the bright unidentified TeV $\gamma$-ray source \hessLHM\
with the X-ray Imaging Spectrometers onboard the Suzaku satellite.  No
X-ray counterpart was found to a limiting flux of
3.1\EHM{-13}~\fluxHM\ in the 2--10~keV band, which is some 60 times
below the $\gamma$-ray flux in the 1--10~TeV band. This object is
bright in TeV $\gamma$-rays but very dim in the X-ray band, and thus
is one of the best examples in the Galaxy of a "dark particle
accelerator."  We also detected soft thermal emission with $kT
\sim$0.3--0.6~keV near the location of \hessLHM.  This may be due to
the dust grain scattering halo from the nearby bright supernova
remnant RCW~103.

\end{abstract}

\section{Introduction}

Since the discovery of cosmic rays, the origin of these high
energy particles has been a mystery. Recent observational
studies have revealed that some supernova remnants (SNRs)
have non-thermal X-ray emission
(e.g. \cite{Koyama1995,Hwang2002,Bamba2003,Vink2003,Bamba2005}),
which can be interpreted as synchrotron radiation from
electrons with energies approaching $E \sim
10^{14}$~eV. TeV $\gamma$-rays have also been detected
from some of these non-thermal shell-type SNRs.  Such very
high energy (VHE) gamma-rays have been explained by either
(1) Inverse-Compton (IC) upscattering of cosmic microwave
background (or other lower frequency) photons by the same
high energy electrons giving rise to the X-ray synchrotron
emission or (2) the decay of neutral pions that originate in
collisions between high energy protons and dense
interstellar matter
(e.g. \cite{Pannuti2003,Lazendic2004,Katagiri2005}).  Thus
the combination of VHE gamma-ray and non-thermal X-ray
emission is expected to be a key observational feature of
high energy particle accelerators. A high sensitivity survey
of the Galactic plane in VHE $\gamma$-rays has been
conducted by the High Energy Stereoscopic System (HESS)
team, who discovered fourteen new VHE $\gamma$-ray sources
\citep{Aharonian2005,Aharonian2006}. \hessLHM\ (hereafter
\hessHM) is one of the brightest of these new sources and,
furthermore, is spatially extended with an angular diameter
of $\sim$\timeform{16'} in the HESS data.  Three known
objects, the young hard X-ray pulsar PSR~J1617$-$5055, the
SNR RCW~103 (G332.4$-$0.4), and the SNR Kes~32
(G322.4$+$0.1), are in the vicinity of \hessHM\ ($\sim$
\timeform{10'}--\timeform{15'} away from it), but none
provides a convincing identification~\citep{Aharonian2005,Aharonian2006}.

Before the HESS Galactic plane survey, a few extended
$\gamma$-ray objects with no clear counterpart in other
wavebands had been found. These objects have been denoted
``dark particle accelerators''\citep{Ubertini2005}.  They
are of interest since they may be indicating the acceleration of 
nucleons; on the other hand, if the TeV emission comes from
electrons, peculiar conditions such as extremely low
magnetic fields may be necessary, due to the short radiative
lifetimes of high energy electrons~\citep{Yamazaki2006}.
TeV~2032+410, discovered by the High Energy Gamma-Ray
Astronomy (HEGRA) collaboration~\citep{Aharonian2002}, was
the first object of this kind.  In spite of several
multiwavelength follow-up studies, no compelling counterpart
has been found from radio to X-ray
wavelengths~\citep{Mukherjee2003,Butt2003,Butt2006}.
HESS~J1303$-$631 is another example~\citep{Aharonian2004}.
\citet{Mukherjee2003} observed this target with the Chandra
X-ray Observatory for 5~ks, and placed an upper limit of
$<$5.4\EHM{-12}\fluxHM\ to its diffuse hard X-ray flux in
the 2--10~keV band.

\hessHM\ is brighter than both TeV~2032$+$410 and
HESS~J1303$-$631 in the TeV regime (see also
table~\ref{tbl:VHEsrc}) and is therefore a more suitable
case for examining if dark accelerators really exist and for
finding out just how ``dark'' they are.  Since many VHE
$\gamma$-ray objects are X-ray emitters, the X-ray band is
the natural place to search for a counterpart.  To date
there have been no X-ray observations explicitly targeting
\hessHM, so we observed \hessHM\ with the Suzaku
satellite~\citep{Mitsuda2006}.  The Suzaku X-ray Imaging
Spectrometers (XIS)~\citep{Koyama2006}, with their high
sensitivity and stable low background~\citep{Yamaguchi2006},
are optimal devices to search for diffuse X-ray emission.
This is especially the case in the hard X-ray band ($E
\gtrsim 6$~keV), where non-thermal emission dominates over
thermal and the effects of X-ray absorption are minimized.
By detecting or even just setting a sensitive upper limit on
the X-ray emission from \hessHM, it should be possible to
address the question of what types of particles, electrons
or protons, are accelerated in the object, and to constrain
the acceleration mechanism by comparing the X-ray and
$\gamma$-ray emission.  In addition to the Suzaku data we
also used XMM-Newton archival data that partly covered the
\hessHM\ region.  In this work, uncertainties are quoted at
the 90\% confidence level unless otherwise stated.

\section{Observations and Data Reduction}

\begin{figure}
  \begin{center}
    \FigureFile(0.5\textwidth,){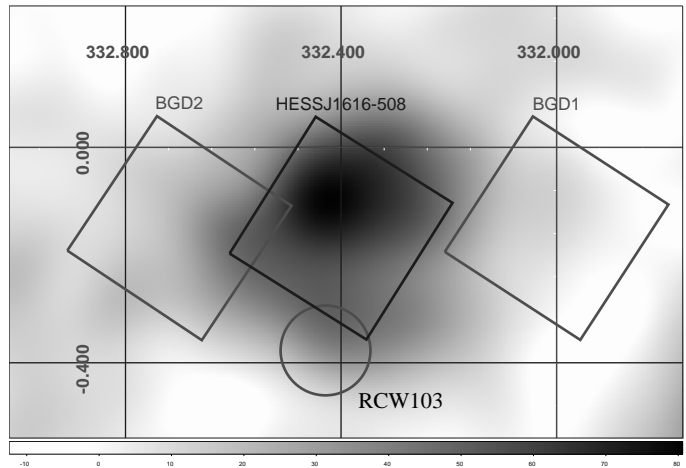}
  \end{center}
  \caption{Suzaku field of view overlaid on the HESS
smoothed excess map in Galactic coordinates (axis labels are
in degrees).  The circle shows the position and approximate
size of RCW~103, which is the nearest X-ray bright SNR. The
scale bar below the figure shows the excess.  }
\label{fig:suzaku_fov}
\end{figure}

The center of \hessHM\ was observed on 2005 September 19
during the Science Working Group (SWG) phase program. We
also observed two nearby positions with the same Galactic
latitude (BGD1 and BGD2 in figure~\ref{fig:suzaku_fov}).
These positions were chosen so that no known bright X-ray
sources fell in the field of view and were close enough to
\hessHM\ that the brightness of the Galactic ridge X-ray
emission
(e.g. \cite{Worral1982,Koyama1986,Yamauchi1993,Sugizaki2001,Revni2006})
could be estimated accurately at the source position.  The
observations are summarized in table~\ref{tbl:obs_log}.

The observations were made with four CCD cameras (designated
as XIS[0-3]; \cite{Koyama2006}) at the foci of four X-ray
telescopes (XRT; \cite{Serlemitsos2006}), and a hard X-ray
detector (HXD; \cite{Kokubun2006,Takahashi2006}). Since the
hard X-ray source PSR~J1617$-$5055~\citep{Torii1998} is in
the HXD field of view, we concentrate on the XIS data.

The XIS was operated in normal clocking mode during all
three observations. The edit mode was
$3\times3$ or $5\times5$ for the \hessHM\ field and we
combined the data of both modes for our analysis, while only
the $3\times3$ mode was used in the background pointings.
We used the HEADAS software version 6.0.4 and version 0.7 of
the processed data
\footnote{Version 0.7 processing is an internal processing
applied to the Suzaku data obtained during the SWG phase,
for the purpose of establishing the detector calibration as
quickly as possible. Official data distributed to guest
observers are processed Version 1.0 or higher. See
\citet{Mitsuda2006} and \citet{Fujimoto2006} for more
details.}. Data affected by the South Atlantic Anomaly,
Earth occultation, and telemetry saturation were excluded.
Hot and flickering pixels were also removed. We did not
screen the data with elevation angle from the bright Earth
in order to maximize the statistics in the hard X-ray band;
more than 99\% of the contamination from the bright Earth
consists of the O~K$\alpha$ and N~K$\alpha$ lines
originating in the atmosphere, and the data above 0.6~keV
are not affected by the contamination. In the following
analysis we use the XIS data from 0.6~keV to 12 keV.  After
data screening, the effective exposures were 45~ks, 21~ks,
and 24~ks for the \hessHM, BGD1, and BGD2 regions,
respectively.

\begin{table*}
\begin{center}
\caption {Log of Suzaku observations.
\label{tbl:obs_log}
}
\begin{tabular}{ccccccccc} \hline \hline
Name   &\multicolumn{2}{c}{Position (degree)}   & Start Time   & End Time   & \multicolumn{4}{c}{Effective Exposure (ks)}\\ 
       & $l$   & $b$  &&&XIS0&XIS1&XIS2&XIS3\\ \hline
\hessHM &332.400  &-0.150  &2005/9/19 12:09:58  &2005/9/20 19:37:50
&45.0&45.4&45.0&45.0\\
BGD1    &332.000  &-0.150  &2005/9/18 22:56:14  &2005/9/19 10:59:02
&21.1&23.7&21.2&21.7\\
BGD2    &332.700  &-0.150  &2005/9/20 19:37:50  &2005/9/21 07:28:46
&23.5&23.7&23.5&23.5\\ \hline
\end{tabular}
\end{center}
\end{table*}

\section{Analysis and Results}

\subsection{XIS image}

\begin{figure*}
  \begin{center}
\begin{minipage}[t]{0.48\textwidth}
    \FigureFile(\textwidth,){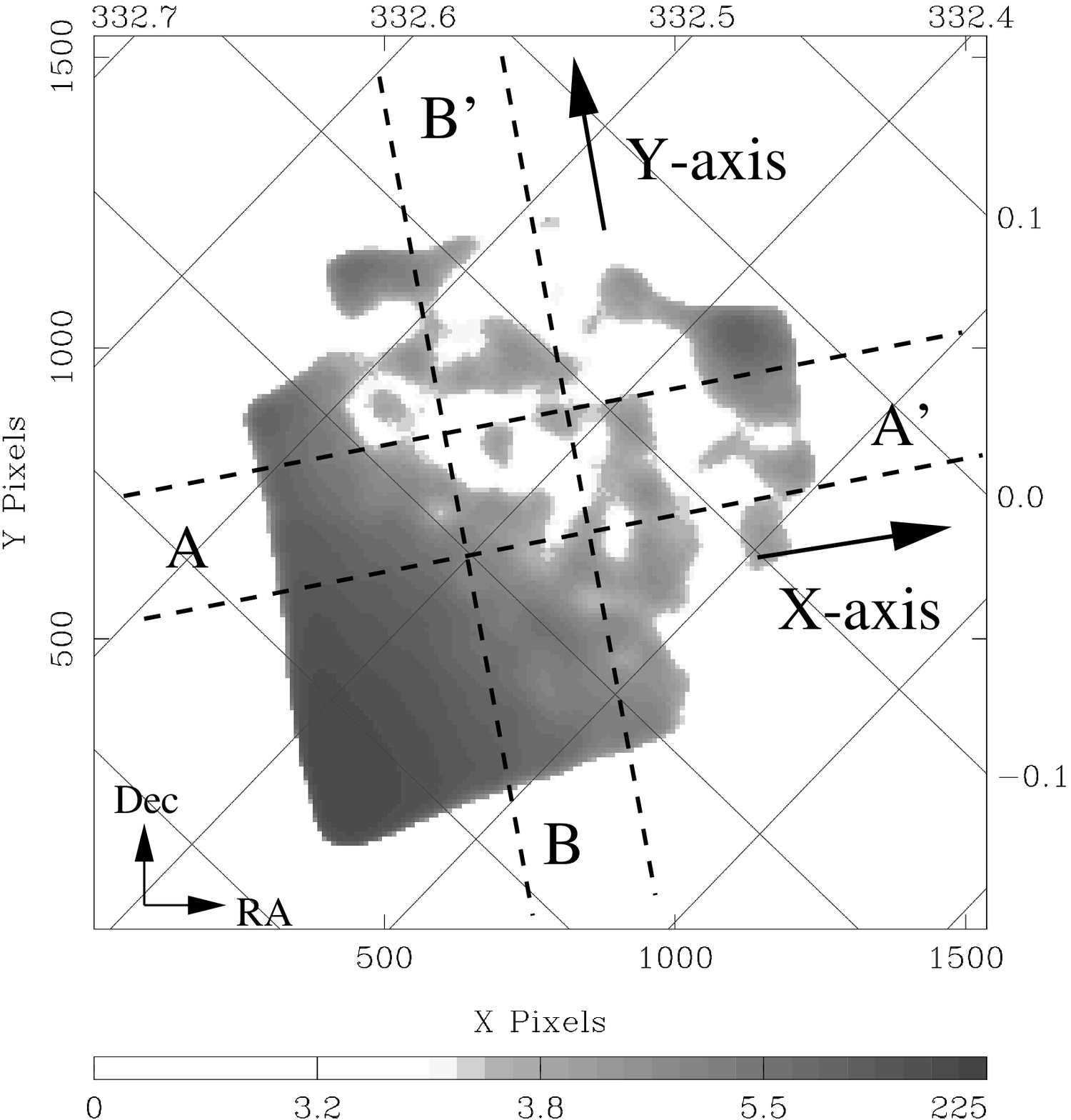}
\end{minipage}
\begin{minipage}[t]{0.48\textwidth}
    \FigureFile(\textwidth,){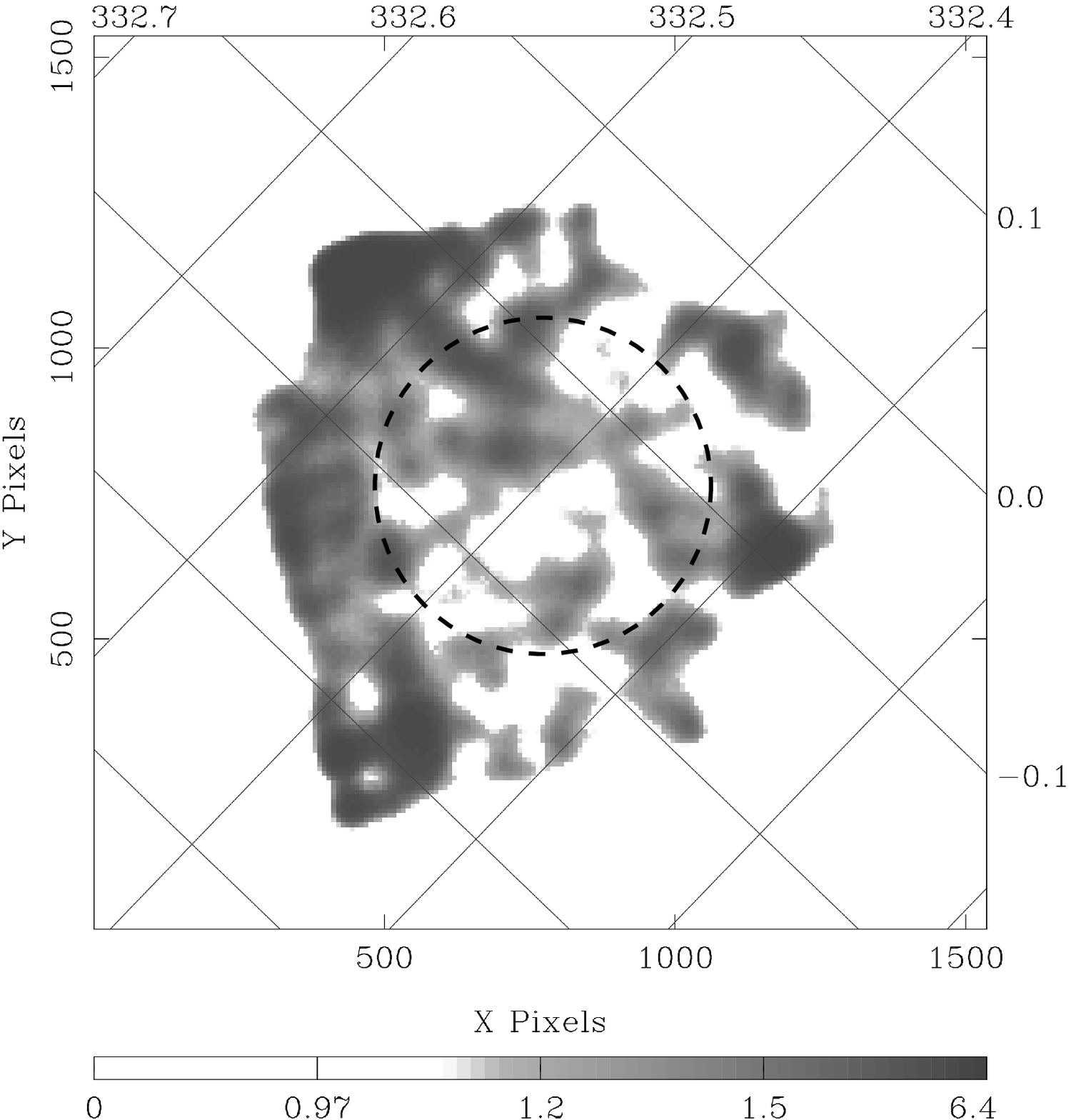}
\end{minipage}
\end{center}
  \caption{Suzaku XIS images of the \hessHM\ field in the
  soft (left; 0.6--3~keV) and hard (right; 3--12~keV) energy
  bands.  Lines of constant Galactic latitude and longitude
  are plotted.  The scale bars under the figures show the
  photon count.  The images were smoothed using a Gauss
  function with a sigma of \timeform{0.42'}. Vignetting
  correction was applied after subtracting non X-ray
  backgrounds, as described in the text.  The dotted lines
  in the left figure show the regions used for the
  photon-count profiles shown in figure~\ref{fig:xis_prof}.
  The dotted circle with a radius of \timeform{5'} in the
  right figure shows the region where we extracted the XIS
  spectra.  }
\label{fig:xis_image}
\end{figure*}

We concentrate on the FI CCD data in this imaging analysis,
since non-X-ray background (NXB) dominates the data of the
BI CCD (XIS1) in the high energy band, especially above
8~keV.  In Figure~\ref{fig:xis_image}, the XIS images of the
\hessHM\ region are shown for the soft (0.6--3.0~keV) and
hard (3.0--12.0~keV) energy bands.  We excluded the corners
of the chips that are illuminated by the $^{55}$Fe
calibration sources, and images from the three FI CCDs
(XIS0, XIS2, and XIS3) were summed.  Images of the NXB were
constructed from night Earth data provided by the XIS team
and subtracted from the \hessHM\ images. After the NXB
subtraction, vignetting corrections were done as follows. We
produced simulated XIS images assuming a uniform surface
brightness, using the XRT+XIS simulator {\tt xissim}
ver.~2006-05-28. We assumed monochromatic X-rays of energy 1.49 or
8.05~keV, where the vignetting function is best
calibrated~\citep{Serlemitsos2006}.  The simulated images
were normalized so that pixels at the optical axis had
values of $\sim$1. We divided the soft and hard \hessHM\
images by the simulated 1.49 and 8.05~keV images.  Finally
the resulted images were rebinned by a factor of 8 and
smoothed using a Gaussian function with a sigma of
\timeform{0.42'}.

\begin{figure*}
\begin{minipage}[t]{0.48\textwidth}
(a)\\
    \FigureFile(\textwidth,){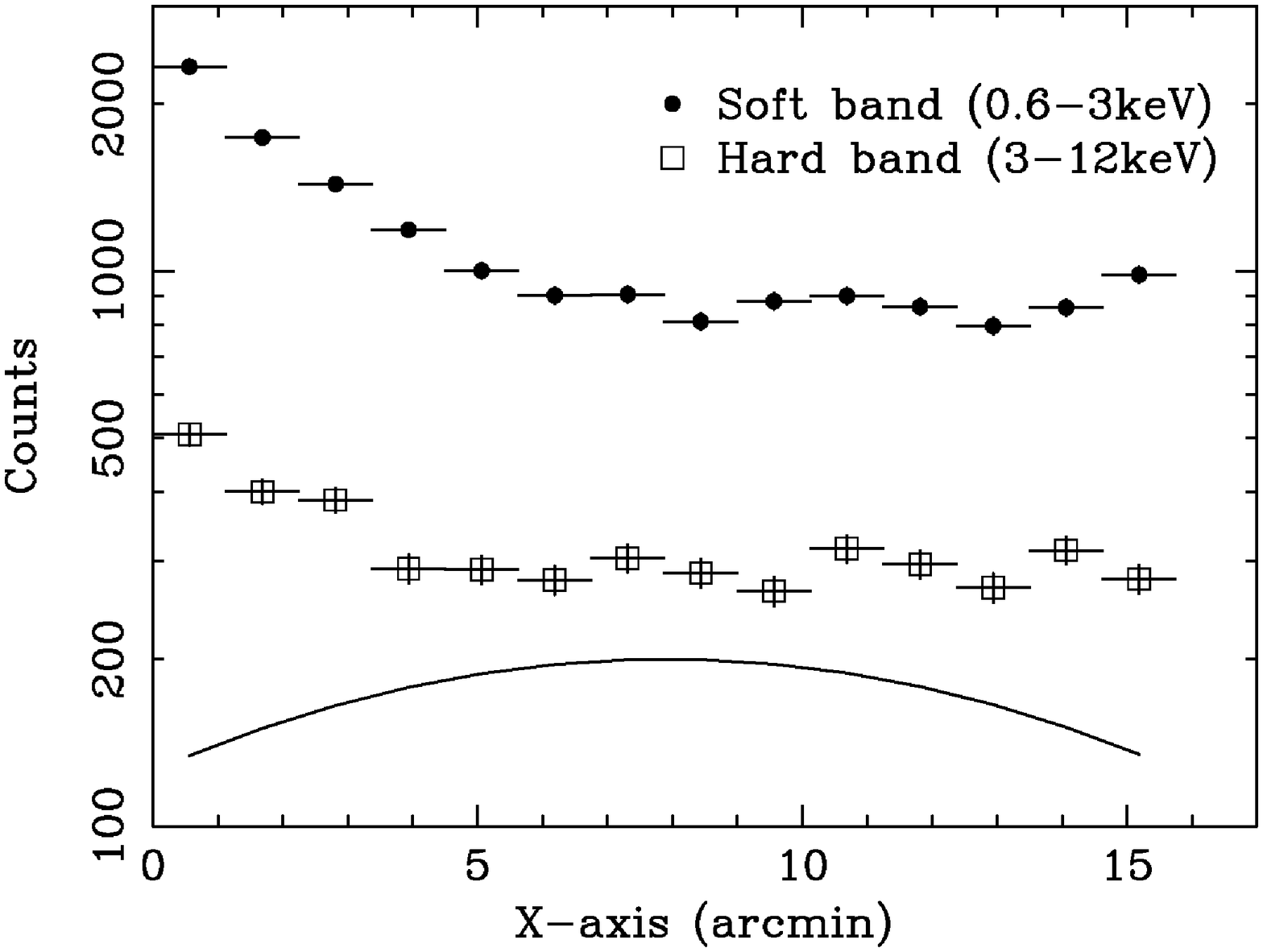}
\end{minipage}
\begin{minipage}[t]{0.48\textwidth}
(b)\\
    \FigureFile(\textwidth,){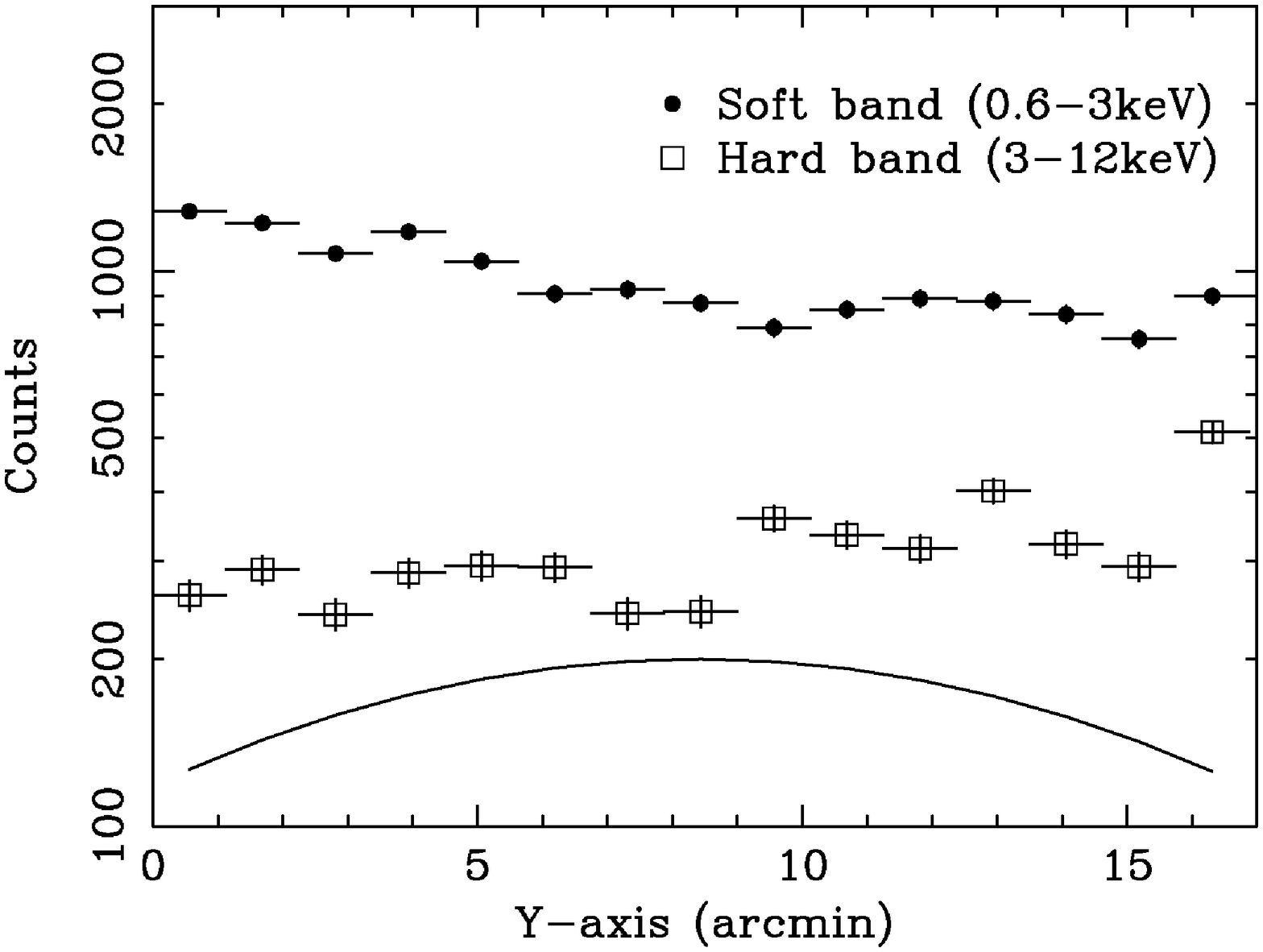}
\end{minipage}

  \caption{Photon count profiles of the XIS images along the
  strips AA' (a) and BB' (b) shown in figure~\ref{fig:xis_image}
The errors are  estimated at the 1$\sigma$ confidence level. 
The curve in the figures shows the Gauss function of $\sigma=\timeform{8.2'}$,
which expresses the TeV~$\gamma$-ray profile of \hessHM.
}
\label{fig:xis_prof}
\end{figure*}

Since the lower left corner of the field of view marginally
overlaps the bright SNR RCW~103, corresponding bright
emission can be seen in the soft band image.  On the other
hand, there is no such structure in the hard band image.
Furthermore, there is no apparent X-ray structure suggesting
an X-ray counterpart of \hessHM\ both in the soft and hard
band images. To examine this more quantitatively, we made
photon count profiles along the strips AA' and BB' with a
width of \timeform{4.2'} in figure~\ref{fig:xis_image}, and
the profiles are shown in figure~\ref{fig:xis_prof}.  

In the soft energy band, the photon count gradually
decreases along both the X and Y axes, which shows the
influence of RCW~103.  A similar trend can be seen even in
the hard energy band along the X axis, but we do not see it
along the Y axis. The trend in the hard energy band may have
a different origin from the soft band profile: 
the hard X-ray pulsar PSR~J1617-5055 or fluctuations of
the galactic ridge emission may explain the hard band
profile.  We see no systematic trend in either the hard or
soft X-ray profiles consistent with the TeV $\gamma$-ray
profile of \hessHM, which can be described by a Gaussian
function with
$\sigma$=\timeform{8.2'}~\citep{Aharonian2006}.  In summary,
according to our imaging analysis, there is no apparent
X-ray counterpart to \hessHM.

\subsection{XIS spectrum}

We made the XIS spectrum of the \hessHM\ region from each
XIS sensor by extracting X-ray events from within a \timeform{5'} radius
of the center of the field of view.  We also tried to
extract spectra for the BGD1 and BGD2 regions in the same
way as for the \hessHM\ region, but we discovered several X-ray
sources in the extraction regions; there is an X-ray object
in the BGD1 region (designated Suzaku~J1614-5114) at $(l,
b)$ = (\timeform{331.98D},\timeform{-0.22D}), and two X-ray
sources are found in the BGD2 region at $(l, b)$ =
(\timeform{332.67D},\timeform{-0.19D}) and
(\timeform{332.77D},\timeform{-0.20D}) (designated 
Suzaku~J1617-5044 and Suzaku~J1618-5040, respectively).
According to the SIMBAD Astronomical Database operated at
CDS, Strasbourg,
France\footnote{http://simbad.u-strasbg.fr/Simbad/},
Suzaku~J1614-5114 and Suzaku~J1618-5040 are positionally coincident
with the B-type star CD-50~10270 and the infrared source
IRAS~16145-5033, respectively. We were unable to identify the
counterpart to Suzaku~J1617-5044. We excluded these sources
from the BGD1 and BGD2 spectra
using \timeform{2'} radius circular regions.  The spectra from 
the FI CCDs were combined after extraction.

\begin{figure*}
\begin{center}
\begin{minipage}[t]{0.8\textwidth}
(a)
    \FigureFile(\textwidth,){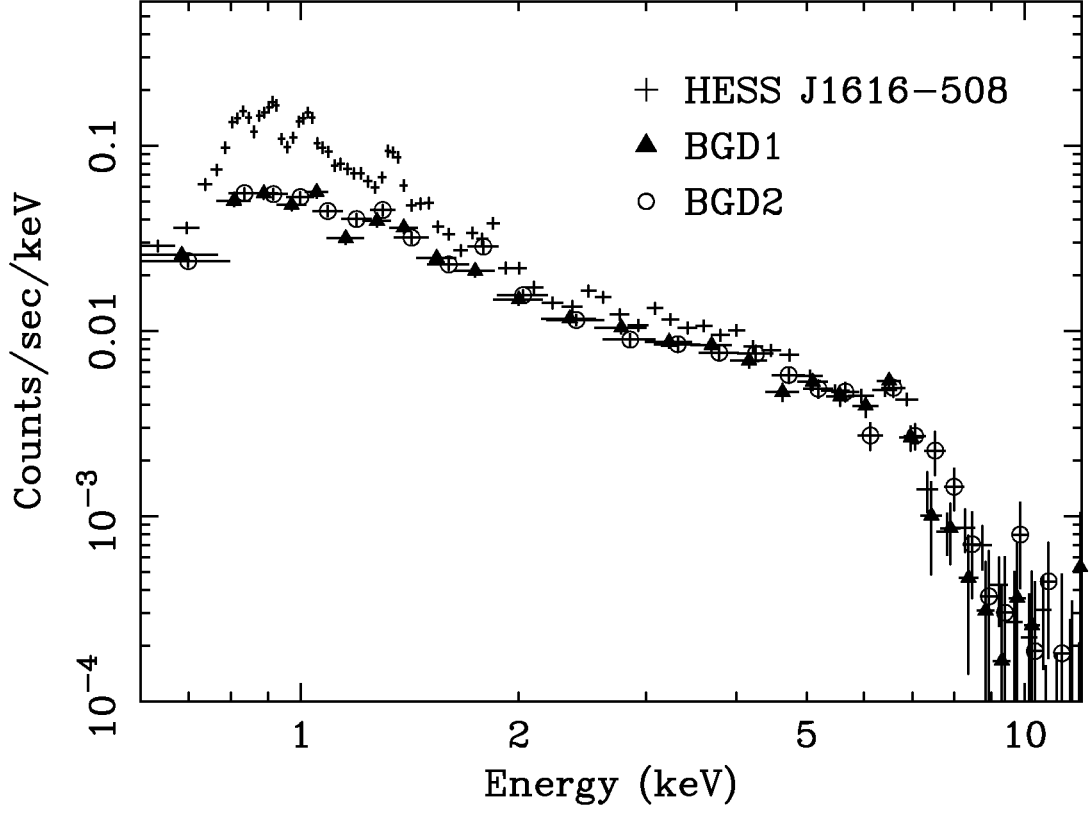}
\end{minipage}
\begin{minipage}[t]{0.8\textwidth}
(b)
    \FigureFile(\textwidth,){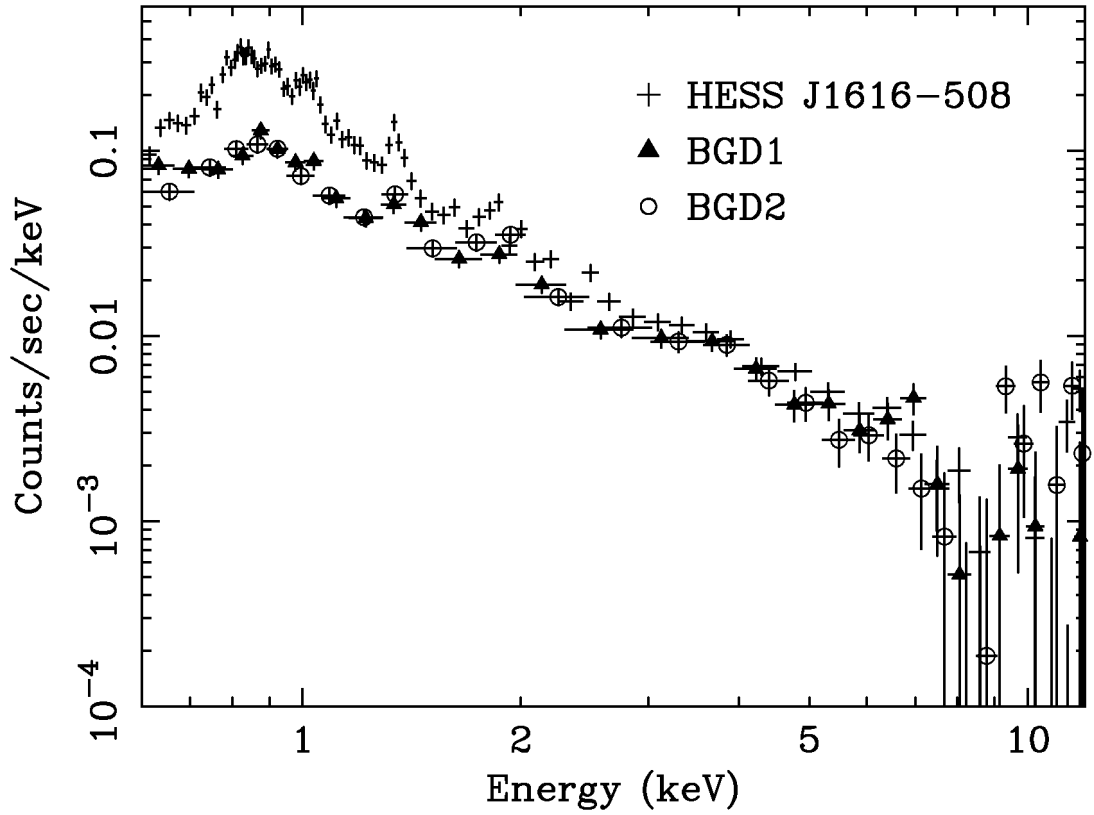}
\end{minipage}
\end{center}
  \caption{XIS spectra from the \hessHM, BGD1 and BGD2
regions: (a) combined FI spectra (XIS0+XIS2+XIS3), and (b)
BI spectra (XIS1).  Non X-ray backgrounds were subtracted as
described in the text. Error bars on the data points are
plotted at the 1$\sigma$ confidence level.}
\label{fig:xis_spec}
\end{figure*}

These spectra contain the NXB.  For the most accurate NXB
estimate, we filtered the night Earth data so that the
cut-off rigidity distribution was the same for the \hessHM,
BGD1 and BGD2 spectra~\citep{Koyama2006}, and extracted the NXB
spectra using the same regions in detector coordinates
(DETX/Y). The NXB spectra thus made were subtracted from the
\hessHM, BGD1, and BGD2 spectra, and the resulting spectra are plotted in
figure~\ref{fig:xis_spec}, where the BGD1 and BGD2 spectra
are renormalized to compensate for the difference in the area of
the spectral extraction regions.

\begin{figure}
\begin{center}
    \FigureFile(.45\textwidth,){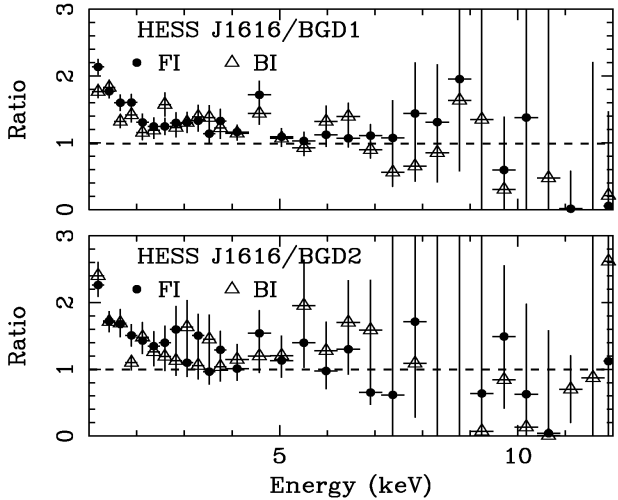}
\end{center}
\caption{Ratio of the spectra shown in
figure~\ref{fig:xis_spec}. Errors on the data points are
1$\sigma$.}
\label{fig:spec_ratio}
\end{figure}

Figure~\ref{fig:spec_ratio} is the ratio of the spectrum of
the \hessHM\ region to those of the background regions.  The
ratio is larger than unity below 4 keV, while it is close to
unity in the 4--8~keV band. The ratio becomes noisy above
8~keV, where the NXB dominates the high energy
band.

The most likely origin for the soft band excess in the \hessHM\ 
region is the nearby SNR RCW 103.  Here we consider whether instrumental
effects, i.e., the tail of the point spread function (PSF) and stray 
light of the XRT, might be the cause.
The \hessHM\ region spectrum exhibits emission lines at 0.83~keV,
0.92~keV, 1.02~keV, and 1.35~keV, which are attributed to
Fe~XVII, Ne~IX, Ne~X, and Mg~XI, respectively, and 
strongly indicate thermal plasma emission. This type of soft spectrum
resembles that of RCW~103.  Using {\tt xissim} we simulated
how many photons would fall into the \hessHM\ region from RCW~103,
which has a radius of $\sim$\timeform{5'} and is centered 
\timeform{13'} from the \hessHM\ region.
In the simulation, we employed a
nonequilibrium ionization plasma model with temperature
$kT=0.3$~keV, column density $N_{\rm
H}=7\EHM{21}$~cm$^{-2}$, and ionization parameter
$nt$=6\EHM{3}~cm$^{-3}$~yr for the spectrum of
RCW103~\citep{Gotthelf1997}.  As for the metal abundances, we set
them to 0.5 times the cosmic values~\citep{Anders1989}. The flux of RCW103
was set to 1.8\EHM{-10}\fluxHM\ in the 0.6--2.0~keV
band~\citep{Tuohy1980} and the Chandra ACIS image of RCW103
(from ObsID 123) was used as the
input surface brightness distribution.  The simulation predicted
count rates of 1.6\EHM{-3}~\crHM\ (FI) and 2.4\EHM{-3}~\crHM\ (BI)
in the 0.6--2.0~keV band. The measured FI count rate from the \hessHM\
spectrum is 4.6\EHM{-2}~\crHM\ with an error of $\pm$0.1\EHM{-2}~\crHM\
depending on the background data set used. The
BI count rate is 9.4\EHM{-2}~\crHM\ with the same uncertainty.
These are more than a factor of 10 {\it higher} than predicted from
the instrumental effects alone, suggesting that the observed soft
emission in the \hessHM\ region may have an astrophysical origin.
However, the current calibrations of the PSF tail and
stray light at large off-axis angles have large
systematic errors. A more detailed understanding of the
XRT will be needed before drawing more definitive conclusions 
about the putative
soft emission, and that is beyond the scope of this paper.

Since the spectral ratio becomes noisy above 8~keV, we
examined the ratio of \hessHM\ to BGDs from 4 to 8~keV. The
average of the ratio and its 99\% confidence range is
found to be 1.10\pmHM{+0.11}{-0.11} and
1.07\pmHM{+0.12}{-0.11} in the cases of \hessHM/BGD1 and
\hessHM/BGD2, respectively.  Thus we cannot conclude a
positive detection of hard X-ray emission from \hessHM\ at the
99\% confidence level with the current statistics of the
Suzaku XIS data.

\begin{figure}
\begin{center}
    \FigureFile(.45\textwidth,){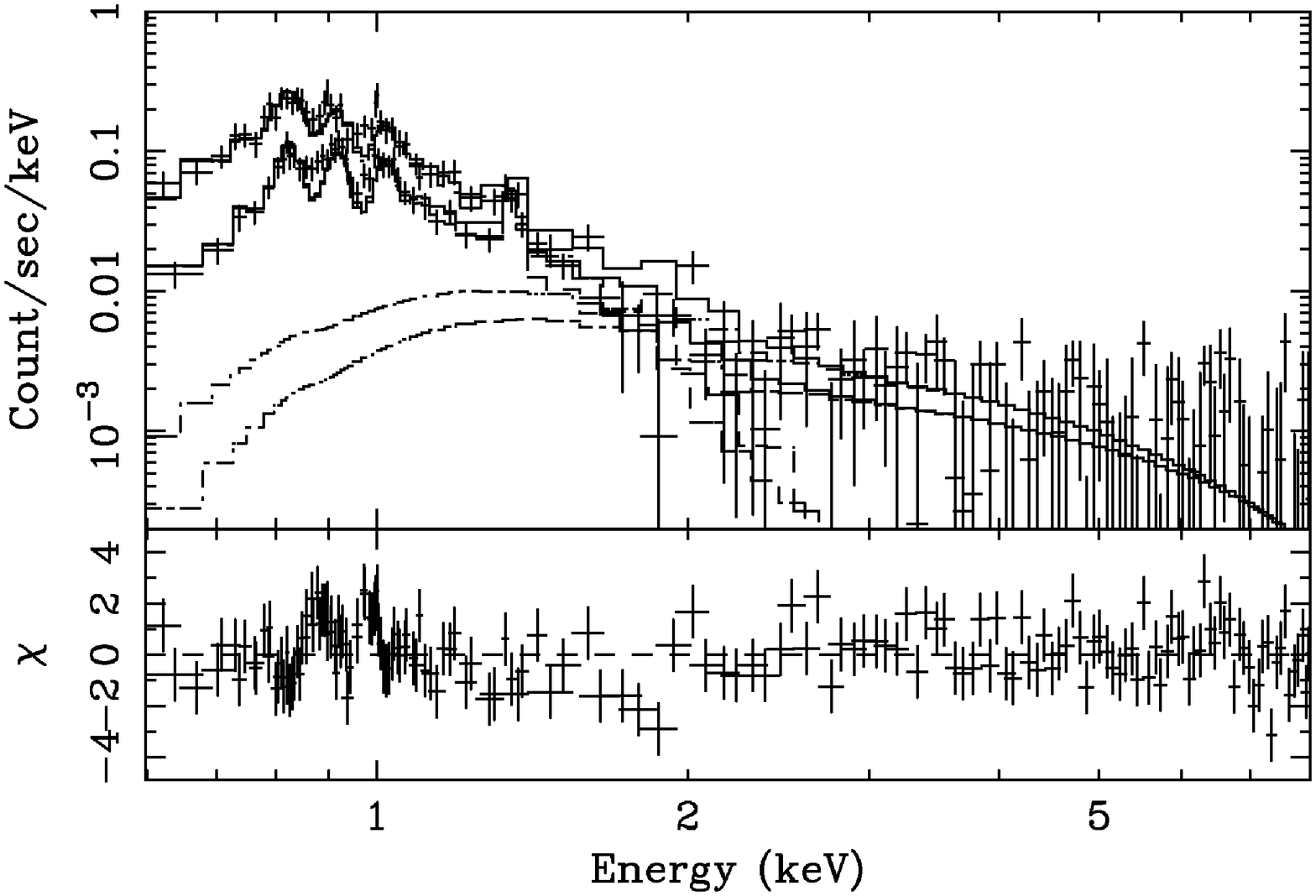}
\end{center}
\caption{\hessHM\ spectra of the FI and BI CCDs with the
best-fit model. The BGD2 data were used for background 
subtraction.  }
\label{fig:bestfit}
\end{figure}

Next we proceed to spectral fitting. Both the FI and BI
spectra of the \hessHM\ region in the 0.6--8~keV band, from
which either BGD1 or BGD2 were used for the background subtraction, were
fitted simultaneously.  To describe the putative soft X-ray
emission, we used a thermal plasma model (the APEC model;
\cite{Smith2001}), and we added a power-law model to quantify
any hard X-ray emission from the \hessHM\ region.  Both components were
modified by the same interstellar absorption.  In the
spectral fitting, we used standard RMFs and we made ARFs for
a flat sky distribution with the software {\tt xissimarfgen}
ver. 2006-05-28. Quoted values for fluxes and
normalizations in the following are normalized to
correspond to a uniform sky distribution over a \timeform{5'} radius.  
The best-fit parameters are summarized in
table~\ref{tbl:results}, and an example of the best-fitted
spectra are shown in figure~\ref{fig:bestfit}.  Note that in
the spectral fits, the power-law component is always
statistically significant even at the 99\% confidence level,
although it is mainly required to fit the
spectra in the 2--4~keV band.  This energy range includes the
soft excess discussed above, which may be related to RCW~103.  
If we replace the power-law model
by a second thermal component, we obtain fits with similar goodness 
of fit. For example, in the
case of the BGD2-subtracted spectra, temperatures of
$kT=$0.62~keV and $kT$=2.0~keV plasma with the same
metallicity of 0.5~times cosmic and $N_{\rm
H}=$3.5\EHM{20}~cm$^{-2}$ can fit the spectra with
$\chi^2/d.o.f$=265.50/237. The 2.0~keV thermal component, which 
may represent contamination from RCW~103 or be part of the putative 
soft emission,  emits 82\% of its energy flux in the 0.6--2~keV band.
Thus the significant detection of a hard power-law
component does not necessarily demonstrate the existence of hard
X-ray emission from \hessHM. Given the uncertainties associated with the
soft diffuse excess we use the power-law component as a
conservative upper limit to the hard X-ray emission.  Furthermore the
BGD2-subtracted case gives the larger value, and so we take
this as the upper limit on the normalization of the power-law
component.  Our upper limit at the 99\% confidence level to the 
unabsorbed flux in the 2--10~keV band of \hessHM\ is
3.1\EHM{-13}\fluxHM.

\begin{table}
\begin{center}
\caption{Best-fit parameters for the model fitting.
\label{tbl:results}
}
\begin{tabular}{lcc}\hline\hline
Background$^a$  &BGD1   &BGD2   \\ \hline
$N_{\rm H}$ ($10^{21}$cm$^{-2}$)&
1.68\pmHM{+0.47}{-0.49} & 4.13\pmHM{+0.45}{-0.45}\\
$kT$ (keV)&
0.59\pmHM{+0.019}{-0.017} & 0.30\pmHM{+0.018}{-0.015}\\
Abundance (cosmic)&
0.50 (fixed) & 0.50 (fixed)\\
$f_{\rm x}^{\rm APEC}$(0.6--2keV)$^b$&
7.05\EHM{-13} & 7.55\EHM{-13}\\
$\Gamma^c$&
2.0(fixed) & 2.0 (fixed) \\
normalization ($10^{-5}$)$^d$&
7.74\pmHM{+2.73}{-2.83} &  9.00\pmHM{+2.88}{-2.92}\\
$\chi^2/d.o.f.$&
273.11/232 & 276.45/232\\ \hline
\end{tabular}
\end{center}

$a$: Data used for background in the spectral fitting.\\
$b$: Flux (\fluxHM) of the APEC model in the 0.6 -- 2~keV band (uncorrected
for absorption).\\
$c$: Photon index of the power-law model.\\
$d$: Normalization of the power-law model defined as the differential 
photon number flux
(i.e., photons~keV$^{-1}$~s$^{-1}$~cm$^{-2}$) at 1~keV. Errors are quoted
at the 99\% confidence level.

\end{table}

\subsection{XMM-Newton analysis}

XMM-Newton~\citep{Aschenbach2000} pointed at the X-ray
pulsar, PSR~J1617$-$5055, on 2001 September 3 (ObsID:
0113050701), and this observation partly covered the
\hessHM\ region with the EPIC instrument \citep{Struder2001, Turner2001}.
We also analyzed the XMM-Newton data to check for
consistency with the Suzaku results.

The pn camera was operated in timing mode during this
observation, and the data have no imaging information. On
the other hand, the MOS1 and MOS2 cameras were operated in
standard full-frame mode using the medium filter.  We
therefore analyzed only the MOS data.  We used the Standard
Analysis System (SAS) software version 6.0.5 for event
selection. From the detected MOS events, we selected those
with PATTERN keywords between 0 and 12 as valid X-ray events. Time
intervals of high and flaring background were rejected by
removing times when the 10--12~keV count rate was higher than
0.15~\crHM\ in the full field of view. The resultant exposure
time was 13~ks for each MOS camera.

\begin{figure}
 \begin{center}
  \FigureFile(0.45\textwidth,){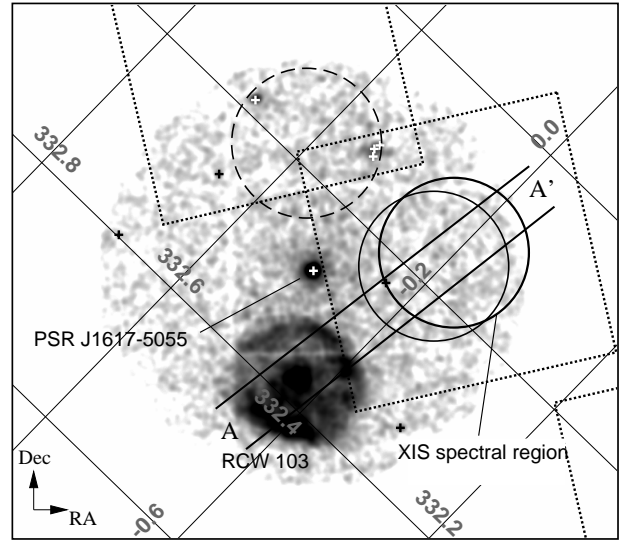}
 \end{center}
\caption{MOS1+2 2.0--7.0~keV image smoothed with a Gaussian
kernel of $\sigma=\timeform{10''}$. Lines of constant
Galactic latitude and longitude are plotted and labeled in
the interior of the figure.  Plotted are the FOVs of the XIS
observations ({\it dashed-line squares}), the XIS spectral
region ({\it thick circle}), detected point sources ({\it
plus marks}), and the source and background spectral regions
({\it solid and dashed thin circles},
respectively). Projection profile in
figure~\ref{fig:xmm_prof} is made along the strip AA'.}
\label{fig:xmm_img}
\end{figure}

Figure~\ref{fig:xmm_img} shows the combined MOS1 and MOS2
image in the 2.0--7.0~keV band.  We see no significant
excess emission at the position of \hessHM.  Because of its
higher spatial resolution, the EPIC observation is
more sensitive to point sources than the XIS
observation even given the shorter exposure time.  So we
searched for X-ray point sources in the combined MOS1+MOS2
2.0--7.0~keV image using {\tt emldetect} in the SAS software
package.
In the {\tt emldetect} software, source significance 
was determined by comparing the numbers of photons in
68\% encircled energy regions to those in the background map, which
was prepared using the {\tt esplinemap} software.  The
likelihood limit was taken to be 10, corresponding roughly
to a 4$\sigma$ detection. The detected point sources, 
excluding those within RCW~103, are shown with plus marks
in figure~\ref{fig:xmm_img}.  Only one point source was
detected at the edge of the XIS spectral region for
\hessHM, but its
flux in the 2--10~keV band is 2.2\EHM{-14}\fluxHM, which is
smaller than the XIS upper limit for the hard X-ray emission
from \hessHM\ by more than a factor of 10.  A source was also
found at around the center of the BGD2 region, but we had
already excluded it from the XIS analysis as
Suzaku~J1617-5044.

The detection limit S$_{lim}$ in the 2--10~keV band of this
search for point sources at the 4$\sigma$ confidence level
was calculated using the formula $S_{lim} \leq (4r\sqrt{\pi
b})/(0.68fT)$, where $b$ is the surface brightness of the
background map, $r$ is the radius of the 68\% encircled
region, $f$ is a counts-to-flux conversion factor (ECF) of
$3.8\EHM{10}$~cts~cm$^{-2}$~erg$^{-1}$, and $T$ is the
``effective'' exposure-time corrected for the vignetting.
For point sources, we obtained S$_{lim}=$2\EHM{-14}~\fluxHM\
at the center of the XMM field of view, while at the center
of the XIS field of view, which is $\sim$\timeform{9'} away
from the optical axis, S$_{lim}=3\EHM{-14}~\fluxHM$.

\begin{figure}
 \begin{center}
  \FigureFile(0.45\textwidth,){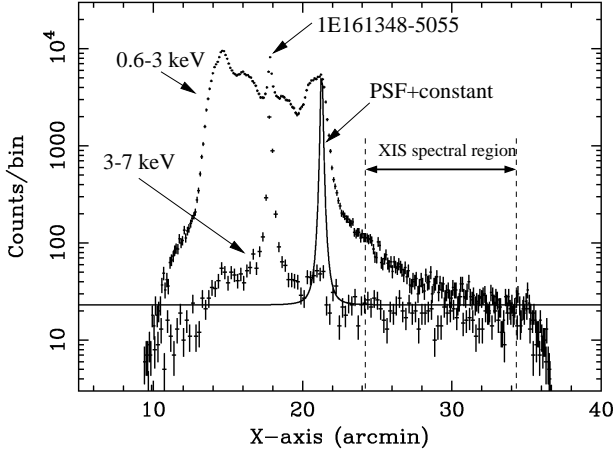}
 \end{center}
\caption{MOS1+2 projection profiles in the 0.6--3~keV and
3--7~keV bands along the strip AA' in
figure~\ref{fig:xmm_img}. }
\label{fig:xmm_prof}
\end{figure}

We made photon count profiles along the strip AA' in
figure~\ref{fig:xmm_img} in the 0.6--3~keV and 3--7~keV band
(figure~\ref{fig:xmm_prof}).  Note that NXB subtraction
and vignetting corrections were not applied to the
profile.  As a reference, we also plot the
PSF\footnote{In XMM-SOC-CAL-TN-0022 at\\
http://xmm.esac.esa.int/external/xmm\_sw\_cal/calib/documentation/index.shtml\#XRT}
at the position of the RCW~103 rim ($\sim$\timeform{6'} away
from the optical axis) plus constant profile in the figure.
The difference between the soft and hard band profiles show 
an extended soft X-ray halo surrounding RCW~103, 
qualitatively supporting the soft excess seen in the Suzaku data.

We also searched for larger scale X-ray
emission at the position of \hessHM\ in the EPIC data.
Source and background spectra are extracted from the
circular regions (\timeform{5'} radii) shown in
figure~\ref{fig:xmm_img} (the solid and dashed thin circles,
respectively).  The background region was chosen to be at
the same galactic latitude as the source region so that 
differences in galactic diffuse emission~\citep{Kaneda1997} 
would be minimized.
The point sources detected in the source and
background region were excluded with \timeform{2''} radius
circles. MOS1 and MOS2 spectra were co-added.  To
avoid the influence of the putative soft emission, we
ignored the data below 3~keV and fitted the spectrum in the
3 -- 7 keV band with a power-law model of a photon index
$\Gamma=2$. The 99\% confidence limit on the power-law
normalization is $<$2.66\EHM{-4}. The XMM-Newton data thus
set an upper limit of 6.9\EHM{-13}~\fluxHM\ on the flux of
the hard X-ray emission from \hessHM\ in the 2--10~keV band.
This upper limit is three times larger than the Suzaku
result.

\section{Discussion}

Neither the Suzaku nor the XMM-Newton data provide evidence
for hard X-ray emission from \hessHM; the XIS sets the most
stringent upper limit of 3.1\EHM{-13}\fluxHM\ in the
2--10~keV band.  The differential photon flux in the TeV
$\gamma$-ray regime can be represented by a power-law of
photon index $\Gamma=2.35$ with a total photon flux of
($43.3\pm2.0$)\EHM{-12}~cm$^{-2}$~s$^{-1}$ above
200~GeV~\citep{Aharonian2006}.  This corresponds to an
energy flux of 1.7\EHM{-11}\fluxHM\ in the 1--10~TeV band.
The ratio of the TeV $\gamma$-ray flux to the X-ray flux
($f_{\rm TeV}/f_{\rm X}$) is therefore more than $\sim$55.
Table~\ref{tbl:VHEsrc} is the list of spatially extended VHE
objects with X-ray observations that we extracted from the
H.E.S.S. source
catalog~\footnote{http://www.mpi-hd.mpg.de/hfm/HESS/public/HESS\_catalog.htm}.
We also added TeV~J2032+4130~\citep{Aharonian2002} to the
list.  Although recent Swift observations found possible
counterparts of HESS~1614$-$518 and
HESS~J1834$-$087~\citep{Landi2006}, we do not include them
in table~\ref{tbl:VHEsrc}, since their X-ray data are not
statistically enough for spectroscopy. Note that the X-ray
couterparts in the list are not secure. For example,
AX~J1838.0$-$0655 is the most promising candidate
counterpart of
HESS~J1837$-$069~\citep{Aharonian2005,Aharonian2006}, it is
slightly outside of the HESS source
extension~\citep{Landi2006}.
\hessHM\ has the most stringent X-ray flux upper limit 
among the VHE objects; yet of more import is its extremely
large flux ratio, $f_{\rm TeV}/f_{\rm X}$, compared to the
others.  Evidently \hessHM\ is a very peculiar object.  The
bright TeV $\gamma$-ray emission strongly suggests the
presence of some particle acceleration processes there;
however \hessHM\ emits little X-ray (or lower frequency
radiation) emission and thus remained undiscovered prior to
the HESS survey. \hessHM\ is thus a strong candidate for
being a ``dark particle accelerator.''

\begin{table*}
\caption{Spatially extended VHE objects with X-ray observations.}
\label{tbl:VHEsrc}
\begin{center}
\small
\begin{tabular}{cccccccccc} \hline \hline
Name   & Possible counterpart &Type$^a$
& $\Gamma_{\rm TeV}^b$ & $f_{\rm TeV}^c$ 
&$N_{\rm H}^d$ & $\Gamma_{\rm X}^e$  & $f_{\rm X}^{f}$ 
&$f_{\rm TeV}/f_{\rm X}$ &Reference$^g$\\ \hline
HESS~J0852$-$463 &RX~J0852$-$4622 &SNR &2.1 &6.9 &4   &2.6 &$\sim$10 &$\sim$ 0.7 &1, 2, 3\\ 
HESS~J1303$-$631 &---           &?   &2.4 &1.0 &20  &2.0 &$<$0.64&$>1.6$ &4, 5\\
HESS~J1514$-$591 &PSR~B1509$-$58  &PWN &2.3 &1.6 &8.6 &2.0 &3.2 &0.5 & 6, 7\\
HESS~J1632$-$478 &AX~J1631.9$-$4752 &HMXB?&2.1&1.7&210&1.6 &1.7 &1.0 &8, 9\\
HESS~J1640$-$465 &G338.3$-$0.0    &SNR &2.4 &0.71 &96 &3.0 &0.30 &2.4 &8, 10\\
HESS~J1713$-$397 &RX~J1713.7$-$3946 &SNR&2.2&3.5 &8 &2.4 &54 &0.065&11, 12\\
HESS~J1804$-$216 &Suzaku~J1804-2142 &?   &2.7 &0.48&2&-0.3&0.025&19&8, 13\\  
HESS~J1804$-$216 &Suzaku~J1804-2140 &?   &2.7 &0.48&110&1.7&0.043&11&8, 13\\  
HESS~J1813$-$178 &AX~J1813$-$178    &?   &2.1 &0.89&110&1.8&0.70&1.3&8, 14\\
HESS~J1837$-$069 &AX~J1838.0$-$0655 &?   &2.3 &1.4 &40 &0.8&1.3&1.1&8, 15\\
TeV~J2032$+$4130 &---             &?   &1.9 &0.20&?  &?  &$<$0.20&$>$1.0&16\\ \hline
HESS~J1616$-$508 &---             &?   &2.4 &1.7&4.1&2.0 &$<$0.031&$>$55&This work\\ \hline
\end{tabular} 
\end{center}
$a$: SNR=supernova remnant, PWN=pulsar wind nebula, HMXB=high mass X-ray binary\\
$b$: Photon index of TeV spectra.\\
$c$: Unabsorbed flux in the 1--10~TeV band (in $10^{-11}$~\fluxHM).\\
$d$: Column density for X-ray spectra (in $10^{21}$~cm$^{-2}$).\\
$e$: Photon index of X-ray spectra.\\
$f$: Unabsorbed flux in the 2--10~keV band (in $10^{-11}$~\fluxHM).\\
$g$: (1)~\citet{Aharonian2005aap437}, (2)~\citet{Slane2001}, 
(3) Our analysis of the ASCA archival data,
(4)~\citet{Aharonian2004}, (5)~\citet{Mukherjee2005},
(6)~\citet{Aharonian2005aap435}, (7)~\citet{Delaney2006},
(8)~\citet{Aharonian2006}, (9)~\citet{Rodriguez2003}, 
(10)~\citet{Sugizaki2001},
(11)~\citet{Aharonian2004Natur}, (12)~\citet{Slane1999}, 
(13)~\citet{Bamba2006}, 
(14)~\citet{Brogan2005}, (15)~\citet{Bamba2003apj}, 
and (16)~\citet{Aharonian2002}
\end{table*}

\begin{figure}
 \begin{center}
  \FigureFile(0.45\textwidth,){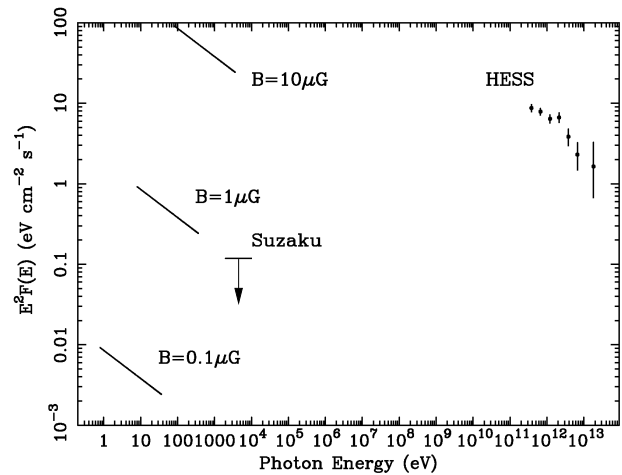}
 \end{center}
\caption{Spectral energy distribution of \hessHM\ from the X-ray to TeV 
$\gamma$-ray band. The synchrotron radiation from accelerated electrons,
which boost the 3~K background up to the TeV energy range, is plotted 
toward the left for three different values of the magnetic field.
}
\label{fig:sed}
\end{figure}

Assuming the origin of the TeV $\gamma$-ray to be IC
scattering of the cosmic microwave background by accelerated
electrons, we can calculate the synchrotron emission from
these electrons assuming magnetic field values of $B=$10, 1,
and 0.1~$\mu$Gauss (figure~\ref{fig:sed}).  The Suzaku flux
limit requires that the magnetic field in \hessHM\ be less
than a few micro Gauss, which is the typical interstellar
value and far below the estimated magnetic field values in
other HESS sources such as RX~J0852$-$4622 and
RX~J1713.7$-$3946. Thus other mechanisms for the TeV
emission may be required. \citet{Aharonian2006}\ proposed an
asymmetric undetected pulsar wind nebula (PWN) powered by
the young pulsar PSR~J1617$-$505, which is
$\sim$\timeform{13'} away from \hessHM, to explain the TeV
emission.  The spin-down luminosity of PSR~J1617$-$505 is
1.6\EHM{37}\lumiHM~\citep{Torii1998}, and in this case the
luminosity of the PWN is expected to be
$\sim$$10^{34}$~\lumiHM~\citep{Cheng2004} in the 2--10~keV
band.  Assuming the pulsar distance to be
3.3~kpc~\citep{Torii1998}, we obtain an unabsorbed flux of
7.6\EHM{-12}~\fluxHM\ in the 2--10~keV band. To reconcile
this high flux and the XIS upper limit, we would need an
absorption of $N_{\rm H}>2\EHM{24}$~cm$^{-2}$ for a photon
index of 2.  Since the column density is more than 10 times
the absorption to the Galactic center~\citet{Baganoff2003},
this seems an unlikely scenario to explain X and TeV
emission.  The high $f_{\rm TeV}/f_{\rm X}$ ratio might be
explained if \hessHM\ is an old SNR colliding with a giant
molecular cloud, since TeV and X-ray emissions are dominated
by hadronic processes and synchrotron radiation from
secondary electrons,
respectively~\citep{Yamazaki2006}. However there are no
radio data indicating the presence of a giant molecular
cloud in the region.  In any case, it remains highly
desirable to obtain detailed data of the \hessHM\ region in
other wavelengths, especially the radio band, to detect the
synchrotron emission from the accelerated electrons.

We found extended soft X-ray emission suggesting a thermal plasma with
$kT\sim0.5$~keV both in the Suzaku and XMM spectra. Both the 
temperature and the projection profiles of the
soft band images (figures~\ref{fig:xis_prof} and
\ref{fig:xmm_prof}) imply a close relationship between the
soft emission and RCW~103.  The best-fit $N_{\rm H}$ of
2--4\EHM{21}~cm$^{-2}$ in table~\ref{tbl:results} also
support the connection, since these values are close to
those obtained with the ASCA
spectra~\citep{Gotthelf1997}. The XMM-Newton profile
(figure~\ref{fig:xmm_prof}) suggests it extends up to
\timeform{15'} from the central point source 1E161348-5055,
which is three times as large as the radius of RCW~103.
Although
we cannot fully reject the possibility, at least for the Suzaku data, 
that this soft emission can be explained by an instrumental effect,
the presence of this emission in the XMM-Newton data argue strongly
for an astrophysical origin.

Many bright X-ray sources in the Galactic plane have
spatially extended X-ray halos, caused by the scattering of 
X-rays into our line of sight by 
interstellar dust between  us and the 
object~\citep{Predehl1995}. The fact that the hard band profile
of XMM-Newton does not extend beyond the rim of RCW~103 may
support the dust halo hypothesis, since the scattering cross
section ($\sigma_{\rm dust}$) depends strongly on photon energy
as $\sigma_{\rm dust} \propto E^{-2}$~\citep{Predehl1995}.
We will address the soft emission again in future when the
calibrations have become much robust.  With careful work it may be
possible to model the soft extended contamination over the
region of \hessHM\ and further reduce the hard band X-ray flux
limit.

\section{Summary}

We observed the bright TeV $\gamma$-ray object, \hessHM,
with the Suzaku XIS for 45~ks. There is no positive
detection of hard X-ray emission and we set an upper limit
of 3.1\EHM{-13}\fluxHM\ to the 2--10~keV band flux.  We also
analyzed the XMM-Newton data of a 13~ks observation and
obtained a three-times higher upper limit of 6.9\EHM{-13}\fluxHM.
The deeper exposure and low background performance of Suzaku explain 
this difference.  The unusually high value of $f_{\rm
TeV}/f_{\rm X}>55$ makes \hessHM\ a very peculiar object.
We also found diffuse X-rays consistent with
thermal emission at $kT\sim$0.3--0.6~keV extending 
over this region from the SNR RCW~103.  This
emission is likely the result of scattering of X-rays
from RCW~103 by interstellar dust.

\bigskip
The authors are grateful to Profs.\ W.~Hoffman and S.~Funk
for kindly providing us the HESS image. We thank
Prof.\ H.~Kunieda for his useful comments.  We also thank all
Suzaku members.  This work is supported by the Grant-in-Aid
for the 21st Century COE "Center for Diversity and
Universality in Physics" from the Ministry of Education,
Culture, Sports, Science and Technology (MEXT) of Japan.  HM
is also supported by the MEXT, Grant-in-Aid for Young
Scientists~(B), 1874015, 2006. JPH acknowledges support from
NASA grant NNG05GP87G.

\end{document}